\documentclass{aa}  
\usepackage{graphicx}
\usepackage{txfonts}
\usepackage{hyperref}
\usepackage{url}
\newcommand{\HI}{H\,{\sc i}}

\newcommand{\kms}{~km\,s$^{-1}$}
\newcommand{\kkms}{km\,s$^{-1}$}

\newcommand{\FHI}{$F_{\rm HI}$}

\newcommand{\MHI}{$M_{\rm HI}$}

\newcommand{\Msun}{~M$_{\odot}$}

\begin{document} 

   \title{The Fourcade-Figueroa galaxy: a clearly disrupted superthin edge-on galaxy\thanks{The reduced  ATCA+GMRT data cube as a FITS file is only available at the CDS via anonymous ftp to
    \url{cdsarc.u-strasbg.fr} (130.79.128.5)
or via \url{http://cdsweb.u-strasbg.fr/cgi-bin/qcat?J/A+A/}}}
   \author{J. Saponara\inst{1,2} \and P. Kamphuis\inst{3}  \and B. S. Koribalski\inst{4,5} \and P. Benaglia\inst{1}
   }

  \institute{Instituto Argentino de Radioastronom\'ia, CONICET-CICPBA-UNLP, CC5 (1897) Villa Elisa, Prov. de Buenos Aires, Argentina
         \email{jsaponara@iar-conicet.gov.ar}
   \and Facultad de Ciencias Astron\'{o}micas y Geof\'{\i}sicas, UNLP, Paseo del Bosque s/n, 1900, La Plata, Argentina    
    \and
    Ruhr University Bochum, Faculty of Physics and Astronomy, Astronomical Institute, 44780 Bochum, Germany
   \and CSIRO Astronomy and Space Science, Australia Telescope National Facility, P.O. Box 76, Epping, NSW 1710, Australia 
   \and 
   Western Sydney University, Locked Bag 1797, Penrith, NSW 2751, Australia  }

  \abstract
  % context heading (optional)
   {Studies of the stellar and the \HI\ gas kinematics in dwarf and Low Surface Brightness (LSB) galaxies are essential for deriving constraints on their dark matter distribution. Moreover, a key component to unveil 
   in the evolution of LSBs is why some of them can be classified as superthin.}
  % aims heading (mandatory)
   {We aim to investigate the nature of the proto-typical superthin galaxy Fourcade-Figueroa (FF),  to understand the role played by the dark matter halo in forming its superthin shape and to investigate the mechanism that explains the observed disruption in the approaching side of the galaxy.}
  % methods heading (mandatory)
   {Combining new \HI\ 21-cm observations obtained with the Giant Metrewave Radio Telescope with archival data from the Australia Telescope Compact Array we were able to obtain sensitive \HI\ observations of the FF galaxy. These data were modelled with a 3D tilted ring model in order to derive the rotation curve and surface brightness density of the neutral hydrogen. We subsequently used this model, combined with a stellar profile from the literature, to derive the radial distribution of the dark matter in the FF galaxy. Additionally, we used a more direct measurement of the vertical \HI\ gas distribution as a function of the galactocentric radius to determine the flaring of the gas disk}
  % results heading (mandatory)
   { For the FF galaxy the Navarro-Frenk-White dark matter distribution provides the best fit to the observed rotation curve. However, the differences with a pseudo-isothermal halo are small. Both models indicate that the core of the dark matter halo is compact. Even though the FF galaxy classifies as superthin, the gas thickness about the galactic centre exhibits a steep flaring of the gas which is in agreement with the edge of the stellar disk. Besides, FF is clearly disrupted towards its north-west-side, clearly observed at both, optical and \HI\ wavelengths. As suggested previously in the literature, the compact dark matter halo might be the main responsible for the superthin structure of the stellar disk in FF. This idea is strengthened through the detection of the mentioned disruption; the fact that the galaxy is disturbed also seems to support the idea that it is not isolation that cause its superthin structure.}


   \maketitle

%-------------------------------------------------------------------

\section{Introduction}

In 1993, Karachentsev, Karachentseva and Parnovskij published the Flat Galaxy Catalogue \cite[FGC,][]{K1993}, and a revised version of it in 1999 \cite[RFGC,][]{K1999}. The catalogue contains disk-like edge-on galaxies with a major-to-minor stellar axis ratio $a/b>7$. 
Superthin galaxies are a special kind of flat galaxies with  $a/b\gtrsim 10$. These are gas-rich Low Surface Brightness (LSB) galaxies, with little or no obvious bulge component, minimal dust absorption \citep{Matthews-2001}, blue optical colours \citep{Dalcanton-2000}, low metallicities \citep{Roennback-1995}, low current star formation and a high ratio of dynamic to \HI\ mass \citep{deBlock-2002}. These characteristics suggest that such galaxies are some of the least evolved galaxies in the Universe. Besides, their high-inclinations and simple structure allow us to study the influences produced by internal as well as external processes \citep{uson-2003}. All the mentioned characteristics set the superthin galaxies as ideal laboratories to investigate the early stages of disk galaxy evolution.

The cosmological paradigm of hierarchical galaxy formation and evolution proposes that galaxies are subject to merging and interaction. Thus, the disk structure and thickness in galaxies are affected by the environment \citep{Toth-1992,odewahn-1994,reshetnikov-1997,schwarzkopf-2001}. This suggests that a flat disk galaxy must remain isolated to persist as a superthin. However, some of these galaxies are found in groups of galaxies as well as in the field \citep{Kautsch-2009}. Therefore, the question arises on how does the pure disk survive? A possible explanation is the presence of a massive dark matter (DM) halo which stabilises their disks against perturbations \citep{1991SvAL...17..374Z,gerritsen-1999}; moreover, \cite{mosenkov-2010} found a correlation between the thickness of stellar disks and relative mass of the dark matter halo. \cite{Banerjee-2013} showed that the determination of the superthin disk distribution, in low-luminosity bulge-less galaxies, is ruled by the compactness of the DM. This idea is supported by the studies performed over the four superthin and proto-typical superthin galaxies  UGC\,7321 \citep{obrien-2010,banerjee-2017}, IC\,2233, IC\,5249 \citep{banerjee-2017} and FGC\,1540 \citep{kurapati-2018}. As this sample is still extremely small any single addition to it can still provide significant new insights or strengthen the current ideas about the nature of superthin galaxies.

Studies of the stellar and the \HI\ gas kinematics in dwarf and LSB galaxies are essential for deriving constraints on the DM distribution \citep{1978ApJ...225L.107R,1998ApJ...499...41M,2010NewA...15...89B}. Moreover, understanding why some LSBs are superthin can contribute the the overall understanding of LSBs and DM in general.
In this paper we focus on the prototypical superthin Fourcade-Figueroa (FF) galaxy. By combining new Giant Metrewave Radio Telescope (GMRT) data with archival Australia Telescope Compact Array (ATCA) data we obtained a high resolution sensitive \HI\ observation of the galaxy. After modelling the gas distribution we use the derived rotation curve to determine the DM distribution in the FF galaxy. 

\begin{figure*}
\centering
\includegraphics[width=14cm]{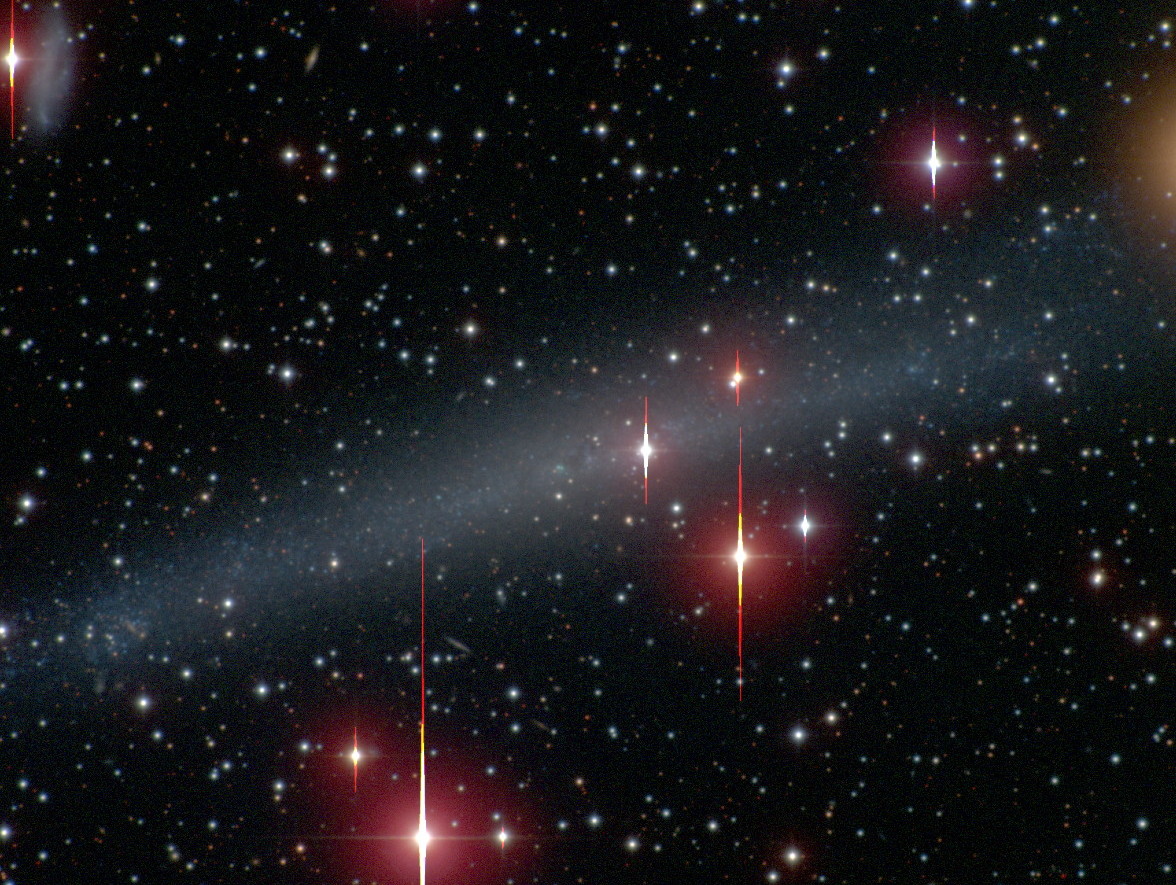}
\caption{FF three-colour image composites from the B, V, and I bands published by \cite{HO-2011}.}
\label{fig:ff-optical}
\end{figure*}

The FF galaxy, also known as ESO\,270$-$G017, was discovered in 1970 \citep{Fourcade} as an elongated and diffuse object, see Fig.~\ref{fig:ff-optical}, located approximately at $2^{\rm o}\,32^{'}$ southeast of the core of the radio galaxy Centaurus A (Cen\,A; NGC~5128). The galaxy was initially thought to be part of the Centaurus A group \citep{1984ApL....24..139C,2013AJ....146...86T,Karachentsev_2014}. However, recent distance estimates place it just beyond Centaurus A at a distance of 7~Mpc \citep{2015ApJ...805..144K}. 
 \cite{K1999 reported a stellar axial ratio of $a/b = 9.1$ (or $b/a=0.1)$.}
The ratio $b/a$ is a strong parameter to determine the flatness of a galaxy, since typical $b/a$ measurements of flat galaxies \citep[between 0.04 to 0.14,][]{K1999} are quite different from the usual values found in the literature regarding all types. 
 The major and minor axes can also be used to derive the galaxy inclination $i$ on a first approximation, as $\cos(i)=b/a$.

We listed the main optical properties of the FF galaxy in Table~\ref{tab:op-properties}. From WISE observations \cite{wang-2017} obtained the two-dimensional structural surface brightness decomposition of FF. This profile has a scale length of 4.4~kpc. From their cleaned 3.4\,$\mu$m WISE image we determine an inner ($r$<2.8~kpc) scale height of 0.47~kpc. This results in a ratio $h_r/h_z$ $\sim$9.4 
 confirming again that the FF galaxy can be considered a superthin galaxy \citep[see also][]{kregel-2002}. FF is a good candidate to perform the mass-modelling and, thereby, determine its dark matter halo. Despite its superthin structure, the FF galaxy shows an asymmetry towards its northwest side \citep{Fourcade}, like a shred or disruption, observed in both optical and \HI\ images. The origin of this disruption remains unclear.\\

The paper is organised as follows: in Sect.~2 we describe the observations and data reduction, including the combination process between data sets from very different radio interferometers; in Sec.~3 we present the \HI\ distribution and kinematics; in Sect.~4 we describe the mass models; in Sec.~5 we present the mass-modelling results and the discussion; in Sec.~6 the shred and in Sect.~7 the summary.

\begin{table}
\centering
\caption{Optical properties of the Fourcade-Figueroa galaxy.}
\begin{tabular}{l@{~~}c@{~~}c@{~~}}
\hline
\multicolumn{2}{c}{Properties} & Ref.\\   
\hline
\hline 
Morphology type                             &  SB(s)m                      & (1) \\
$\alpha$ (J2000)  [$\rm h \ m \ s$]                     &  13:34:47.3       & \\
$\delta$ (J2000)  [$\rm ^o \ ' \ ''$]                    &  $-$45:32:51      & \\
$D$ [Mpc]                              & 6.95                          & (2)\\
$m_{\rm B}$      &  11.7                        &  (3)\\
$D_{\rm B25}$ ['] & 11.3  & (3) \\
$A_{\rm B}$              & 0.48 & (4)\\
$M_{\rm B}$ [mag]        & -18 & \\
$L_{\rm B}$ [L$_{\odot,B}$]   &2.4$\times$10$^9$& \\
$F_{\rm H_{\alpha}}$ [$\rm erg \ cm^{-2} \ s^{-1}$] &(2.09$\pm$0.37)$\times$10$^{-12}$& (5) \\
$\log(SFR)$ [$\rm M_{\odot} \ yr^{-1}$]    &-1.0& \\
$i$  [deg]        &   83                         & (4)\\
 a/b &9.1& (6) \\
\hline
\end{tabular}
\label{tab:op-properties}
{\flushleft Ref.: references; (1) \cite{RC3}, (2) \cite{2015ApJ...805..144K}, (3) ESO LV  \cite{1989spce.book.....L}, (4) \cite{1998ApJ...500..525S}, (5) \cite{kennicutt-2008}, (6) \cite{K1999}.}
\end{table}

\section{HI Observations and data reduction} 

The FF galaxy has been observed at 1420~MHz with the Australia Telescope Compact Array (ATCA). The data were downloaded from the Australia Telescope Online Archive (ATOA\footnote{\url{https://atoa.atnf.csiro.au/}}). The data were published by \cite{2018MNRAS.tmp..467K} and are part of the Local Volume \HI\ Survey (LVHIS) project. Notwithstanding the existence of previous observations, new ones were carried out at the same frequency with the GMRT, that provides better angular and velocity resolution. Additionally, it has better coverage of the $uv$-plane which can be further improved upon by using observations from both telescopes. In the following subsections we describe the observations and data calibration for both ATCA and GMRT as well as the combination process.

\begin{table*} 
\centering
\caption{ATCA and GMRT observing parameters.}
\begin{tabular}{lcccc}
\hline
&\multicolumn{3}{c}{ATCA configuration} & GMRT\\

                       & 750A     & 6A        & EW367     &            \\
\hline
\hline
Project                & C245     & C245     &  C1341    &   $\rm 28\_069$   \\
Date(s)                   & 28-01-93/29-01-93 & 28-06-93/16-07-93 & 15-11-08  &    29-06, 04-07-15/       \\
                       &  & &           &    07-07-15/26-01-16     \\
Time on source [min]  &  474.9   &  588/544     &  597.0    &     195/501/196          \\
                  
Centre frequency [MHz] &  1417    & 1417     &   1417    &     1414       \\
Bandwidth [MHz]        &   8      &   8      &    8      &     4.16 \\
Number of channels        &  512     &  512     &   512     &     512       \\
Channel width [\kkms]  &  3.3     &  3.3     &   3.3     &     1.8       \\
Velocity resolution [\kkms]  & 4  &  4       &   4       &     1.8       \\
Primary beam   [$'$]               &  \multicolumn{3}{c}{33}                            &     26.6            \\
\hline
\end{tabular}
\label{tab:obs}
\end{table*}

\subsection{Australia Telescope Compact Array data}

The ATCA 21-cm observations were carried out in January 1993, June and July 1993 as well as November 2008 using the 750, 6A and EW367 array configurations respectively. The total integration time on-source was 27~h 40~min. The 8~MHz bandwidth divided into 512 channels resulted in a spectral resolution of 15.6~kHz, equivalent to $\sim$4\kms\ per channel. A flux calibrator was observed at the beginning and the end of each observing run for 10 min while a phase calibrator was observed between the target scans every 45 min; see Table~\ref{tab:obs} for more details. 

Data reduction and analysis were performed with the {\sc miriad} software \citep{Sault1995} using standard procedures. We calibrated each data set separately, using PKS\,0407--658 (750/6A array) and PKS\,1934--638 (EW367 array) as primary flux and bandpass calibrators. PKS\,1320--446 (750/6A array) and PKS\,1421--490 (EW367 array) served as the phase calibrators.
We used the {\sc miriad} task {\sc uvlin} to subtract the continuum. With a bandwidth of 8~MHz (see Table~\ref{tab:obs}), we were able to select the line-free channels on either side of the detected \HI\ emission.

\subsection{Giant Metrewave Radio Telescope data}

We observed the FF galaxy with the GMRT at 21-cm for a total time of $\sim$14~h (project 28\_069, PI P.  Benaglia). The observations were carried out during June-July 2015 and January 2016 in the spectral zoom mode with the GMRT-Software backend. We used a 4.16~MHz bandwidth with 512 channels which results in a spectral resolution of 8.13~kHz, equivalent to 1.7\kms\ per channel. The flux calibrator 3C\,286 was observed at the beginning and the end of the run. The source 1323--448 was used as the phase calibrator and was observed between the target scans; see Table~\ref{tab:obs} for more details. 

The data were flagged and calibrated using the FLAGCAL pipeline \citep{2011ascl.soft12007P}. Besides, we extensively used for further analysis, visualisation and multi-wavelength imaging the {\sc miriad} software package and kvis, part of the karma package \citep{Gooch1996}. To check the calibration process, all the calibrator sources were imaged. GMRT does not do online Doppler tracking. Thus, the Astronomical Imaging Processing System \citep[AIPS,][]{Greisen2003} task CVEL was implemented to apply the Doppler shift corrections. All the data sets were combined using the AIPS task DBCON. The continuum subtraction was made with {\sc miriad} task {\sc uvlin}; we were able to select the line-free channels on either side of the detected \HI\ emission.

\subsection{Combining ATCA and GMRT radio interferometer data}

The benefits of complementing the GMRT with ATCA data are the improvement of the signal-to-noise ratio by $\sim$15\%, a better $uv$ plane coverage that should improve the quality of the final images, and a better angular resolution. In this manner, we will get the most out of observations; we can map the extended structures with excellent angular resolution.\\

As the GMRT is a non-coplanar array we in principle should account for this when imaging the data, i.e a $w$-projection is required. However, for our current observations we estimate the shift introduced by ignoring this effect would lead to a shift of 2.25" at the first null of the primary beam over a 12h observation. This leads to a 10$\%$ error on the synthesised beam if we limit our baselines to be < 28 k$\lambda$. As {\sc miriad} is the only package that allows us to combine the data in he uv-domain while taking into account the primary beam correction we choose to image the data without a $w$-projection. As for both observation the band width was too narrow to perform a self-calibration, we conclude that this problem will not affect the image cubes if we just use the visibilities up to 28~k$\lambda$ (equivalent to 6~km). 

We carried out the data combination in the $uv$ domain using the {\sc miriad} software. Both data sets need to have the same velocity system reference and channel increment organised in the same way;  the tasks CVEL, SPLIT and BLOAT from AIPS were used to arrange the visibilities of the different observations on the same frequency grid. We implemented the {\sc miriad} routine {\sc invert} to perform a linear mosaic of these 21-cm GMRT and ATCA data.  The {\sc invert} task optimises the signal-to-noise ratio using the system temperature. For such, once loaded the GMRT data sets into {\sc miriad} we had to add the value of the system temperature in the header. This parameter was gathered from the GMRT user's manual\footnote{http://gmrt.ncra.tifr.res.in}. The deconvolution process was carried out using the Maximum Entropy deconvolution, {\sc mosmem} task, for a mosaiced image. The final synthesised beam was created with task {\sc mosspsf} and applied during restoration.

 To summarise, the final \HI\ cube was made using baselines up to 28 k$\lambda$ and natural weighting. The synthesised beam is $20''\times 20''$, which results in a physical resolution of 673~pc~$\times$~673~pc at the adopted FF distance. The corresponding r.m.s. noise is 2 mJy\,beam$^{-1}$.

\begin{figure}
\includegraphics[width=0.45\textwidth]{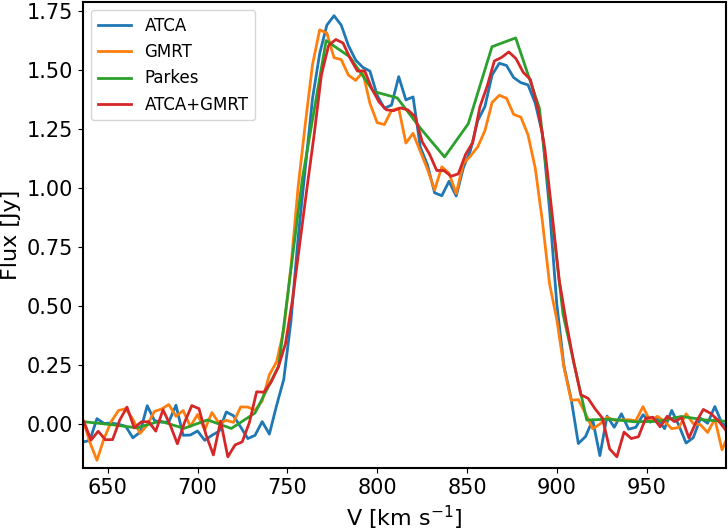}
\caption{ Comparison of the FF global \HI\ profile obtained using ATCA (light blue), GMRT (orange), Parkes \citep[green, ][]{2004AJ....128...16K} and the combination of ATCA and GMRT data (red).}
\label{fig:profile}
\end{figure}

\begin{figure*}
\centering
\includegraphics[width=14cm]{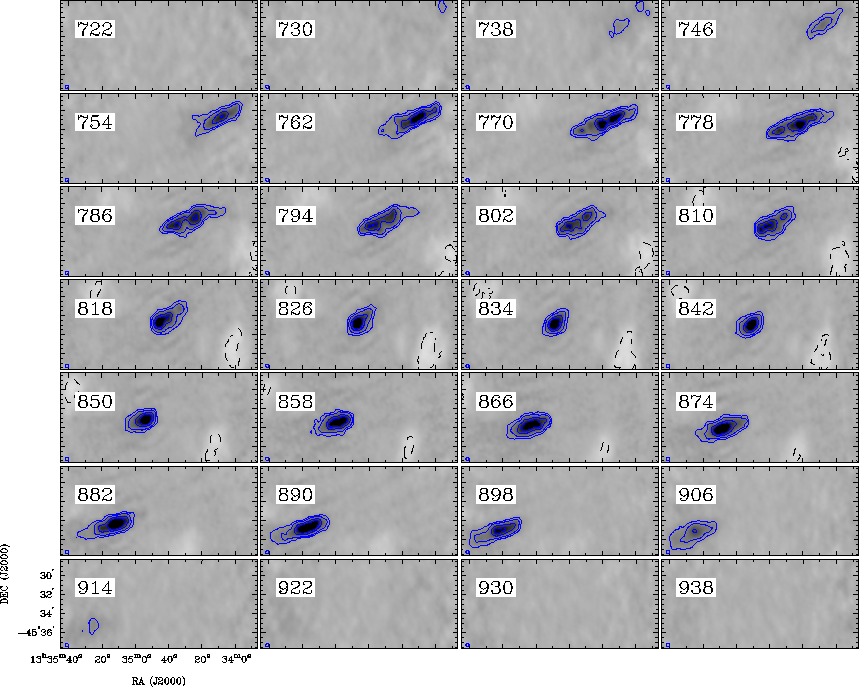}
\caption{ATCA+GMRT high-resolution \HI\ channel maps of the Fourcade-Figueroa galaxy. The channel velocity is shown in the top left corner (in\kms) and the synthesised beam ($20''$) in the bottom left corner of each panel. The contour levels are --6, 6, 15, 25, 35 mJy\,beam$^{-1}$. }
\label{fig:channels}
\end{figure*}

\begin{figure*}
\centering
\includegraphics[width=0.45\textwidth]{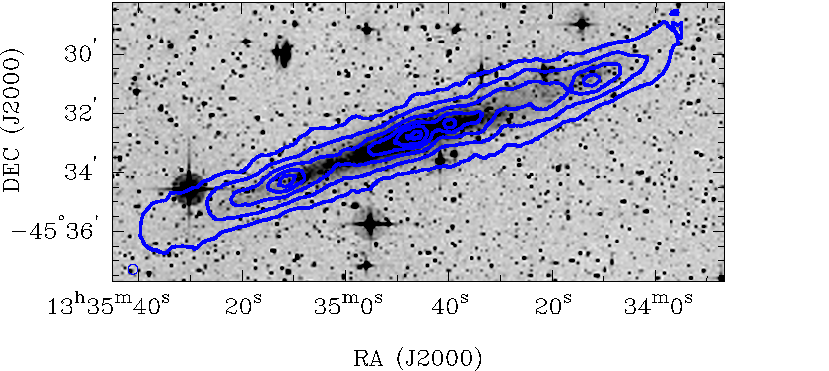}
\includegraphics[width=0.45\textwidth]{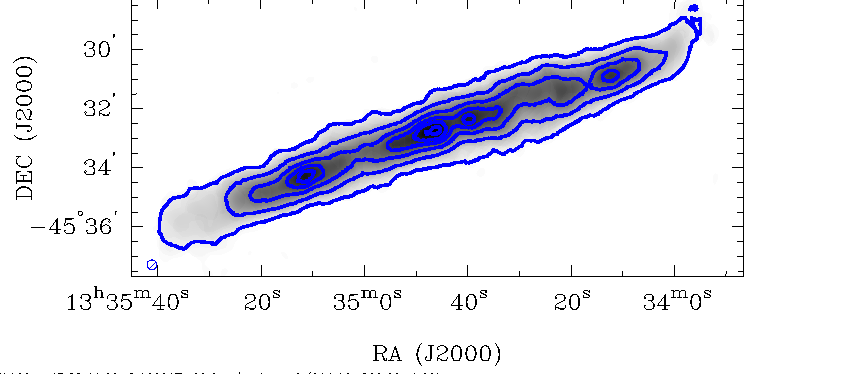}\\
\includegraphics[width=0.45\textwidth]{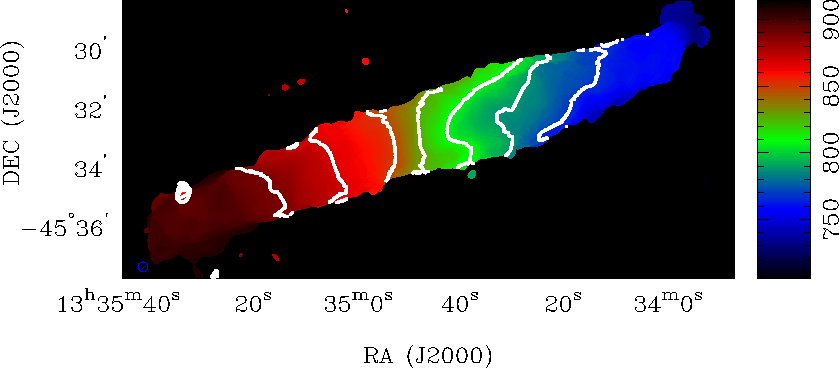}
\includegraphics[width=0.45\textwidth]{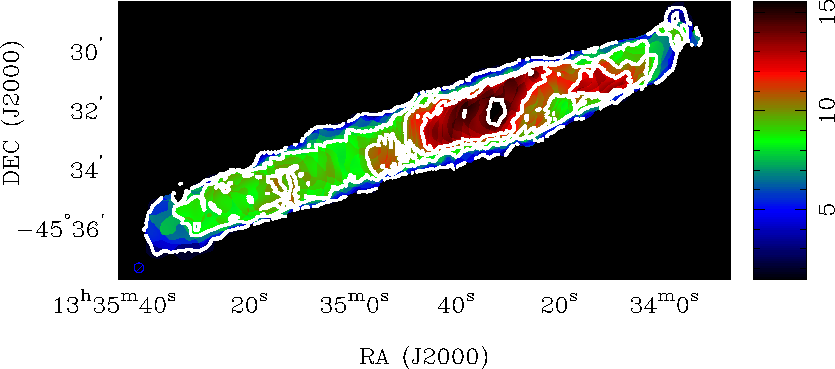}
\caption{ATCA+GMRT \HI\ moment maps of the Fourcade-Figueroa galaxy. {\bf Top left panel:} \HI\ distribution overlaid on the DSS2 $I$-band image. The \HI\ contour levels are 0.24, 0.9, 1.7, 2.1 Jy beam$^{-1}$\kms. {\bf Top right panel:} \HI\ distribution (same contours).  
{\bf Bottom left panel:} velocity field, the contour levels are 768, 788, 808, 828, 848, 868, 888\kms. {\bf Bottom right panel:} velocity dispersion; the contour levels are 3, 8, 10, 12, 15\kms. The \HI\ distribution maps were made using the low-resolution cube with a synthesised beam of $20'' \times 20''$.}
\label{fig:combine-rob-moments}
\end{figure*}

\section{HI distribution and kinematics}

To ensure that our combined data set is free from systematic biases we first extract the line spectrum and compared it to the individual datasets and an archival single dish observation. The results are shown in Fig.~\ref{fig:profile}. The total line flux obtained from the combination data cube is \FHI\ = 197$\pm$0.4~Jy\kms, which is in agreement with the value present in the \HI\ Parkes All-Sky Survey (HIPASS), \FHI\ = 199.4$\pm$15.1~Jy\kms\ \citep{2004AJ....128...16K}, and LVHIS \citep{2018MNRAS.tmp..467K} with the LVHIS \FHI\ = 224.7~Jy\kms. 

The global \HI\ profile of the FF galaxy is shown in Fig.~\ref{fig:profile}.
 It can be appreciated that in the case of the profile derived from the combined cube (ATCA+GMRT data), the flux along the velocity axis remains, mostly, equal or below that derived from Parkes data alone, as it is expected.

The \HI\ diameter is $D_{\rm HI} = 20'$. The \HI\ emission was detected from $\sim$720 to 920\kms. 
At a distance of 6.95~Mpc, this corresponds to a total \HI\ mass of \MHI=2.2$\times 10^9$\Msun, which means that \MHI/$L_{\rm B}$=1.04. The systemic velocity of FF is 828\kms. The profile widths of 20$\%$ and 50$\%$ levels are 142\kms\ and 120\kms.

\subsection{\HI\ distribution}

Fig.~\ref{fig:channels} shows the channels where the FF galaxy contains detectable \HI\ emission. A visible disruption is observed in the channels covering the velocity ranges from 770 to 810\kms.  The shred \citep{Fourcade,1992MNRAS.257..689T} is clearly seen as a disruption in the distribution as well as the kinematics of the gas, and corresponds to the shred as seen in the optical. The disruption is at a projected radial distance $\sim$5~kpc from the centre, same position where the \HI\ distribution bends away from the major axis to the north, see Fig.~\ref{fig:combine-rob-moments}. The \HI\ is distributed over a diameter of $\sim 20'$, almost two times the optical diameter ($\sim 11.3'$), see Fig.~\ref{fig:combine-rob-moments} top panels. The velocity field of FF is not regular as can be seen in Fig.~\ref{fig:combine-rob-moments} bottom left panel. The velocity dispersion varies between 5 and 15\kms, see Fig.~\ref{fig:combine-rob-moments} bottom right panel.  Note that since we are looking at a (nearly) edge-on galaxy, the measured dispersion indicates the spread in coherent rotation along any line of sight, rather than a physical gas velocity dispersion.\\

\subsection{The vertical gas thickness}\label{sec:thickness}

Considering the \HI\ gas distribution in FF is asymmetric, we derive the thickness of the \HI\ disk as a function of the galactocentric radius. The method we implemented follows the procedures laid out in \cite{Olling-1995} and \cite{obrien-2010}. The vertical \HI\ profile of FF exhibits a high degree of asymmetry of the gas thickness about the galactic centre. In the receding side of the galaxy, the \HI\ flares from a Full Width Half Max (FWHM) of 2.5~kpc at 5~kpc out to 3.5~kpc at 18~kpc. The steep flaring of the gas is in agreement with the edge of the stellar disk. Between 5~kpc and 15~kpc, the \HI\ thickness seems to remain constant, except for a bump at ~8~kpc. On the approaching side, the vertical \HI\ thickness shows an irregular flaring profile. From 3~kpc to 5~kpc the \HI\ thickness is roughly constant with an FWHM around 2~kpc. In the range between 5~kpc to 10~kpc, the \HI\ gas flare from 2.3~kpc to 3.6~kpc which is in agreement with the location of the shred. The steep gas flare at the mentioned galactocentric distance is coincident with the radius where the stellar disk becomes fainter. Additionally, the line width of the velocity profiles (see Fig.~\ref{fig:combine-rob-moments}) highly increases at this position. After 10~kpc (galactocentric radius), the \HI\ disk thickness decreases at radii outside of 15~kpc; this is in agreement with the decrease of the \HI\ velocity dispersion.  The linear size of the beam is $\sim700$~pc, and Fig. 5 shows that variations on vertical gas thickness occur on a much larger scale (a few kpc, thus including several beams), then overriding --or at least minimising-- beam smearing problems. On average $z_0$ is 32.6\arcsec$\pm$0.2\arcsec or $z_0$=1.1$\pm$0.4~kpc at the assumed distance of 6.95~Mpc.

\begin{figure}
\centering
\includegraphics[width=0.5\textwidth]{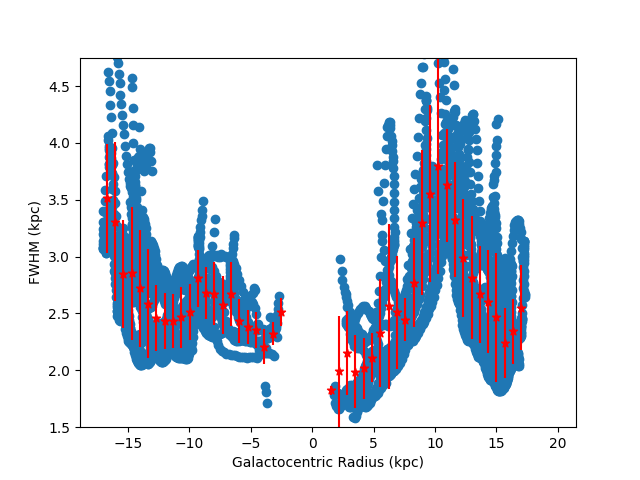}
\caption{Measured \HI\ FWHM thickness (negative radius is receding).}
\label{fig:FF-thickness}
\end{figure}

\subsection{H\,{\sc i} kinematics, 3D modelling and the rotation curve using FAT $\&$ TiRiFiC}

\begin{figure}
\centering
\includegraphics[width=0.43\textwidth]{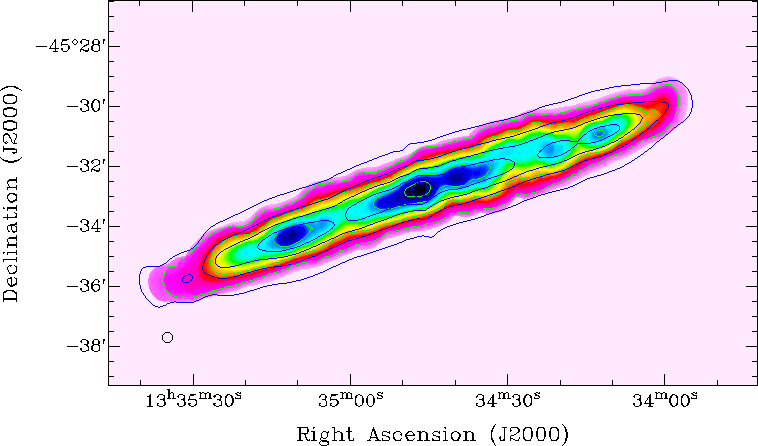}
\includegraphics[width=0.45\textwidth]{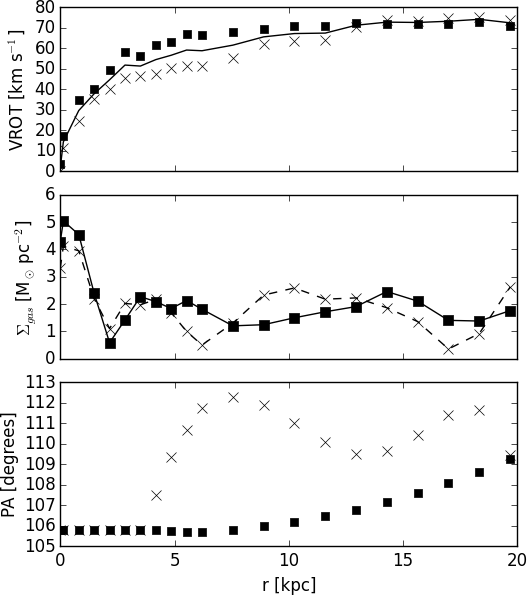}
\caption{{\bf First row}: Comparison between data (green) and model (blue) moment zero maps. {\bf From 2nd to 4th rows}: Model parameters as a function of radius. The crosses/squares represent the approaching/receding halves of the FF galaxy. {\bf Second row}: Rotation curve, the solid curve corresponds to the mean rotation curve. {\bf Third row}: The de-projected \HI\ gas surface distribution. {\bf Fourth row}: The position angle variation. }
\label{fig:ff-rotation}
\end{figure}

\begin{figure}
\centering
\includegraphics[width=0.45\textwidth]{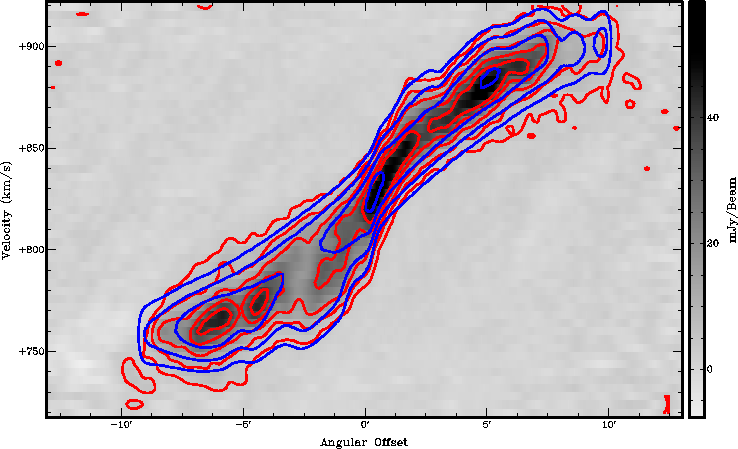}
\caption{The \HI\ position-velocity ($pv$) diagram along the major axis is shown in grey colours and red contours while the blue contour levels represent the $pv$ of the model galaxy; the contour levels are 4, 9, 17, 29, 39 mJy beam$^{-1}$\kms.}
\label{fig:model+data}
\end{figure}

The rotation curve was derived using the Fully Automatic TiRiFiC code \cite[FAT, ][]{2015ascl.soft07011K}, which is a wrapper code around the Tilted Ring Fitting Code \citep[TiRiFiC, ][]{2012ascl.soft08008J}, to perform a 3D modelling, and the Source Finder Application \cite[SoFiA, ][]{Serra-2015} to estimate the initial parameters. As it turns out FAT is not capable of dealing with the varying noise statistics, due to the primary beam correction. The final FAT model did not cover the full extent of the \HI\ disk. Therefore the outer parts of the galaxy were fitted manually with TiRiFiC itself.  Performing a 3D modelling, by fitting the 3D observations directly as TiRiFiC and FAT routines do, guarantees that problems such as beam smearing and projection effects are overcome, according to \cite{2012ascl.soft08008J} and \cite{2015ascl.soft07011K}.
Since the FF galaxy is clearly warped and asymmetric, we created a 2-disk model where each disk represents one half of the galaxy. The inclination was the most problematic parameter to fit; we attempt values between 80 and 90 degrees and after the visual inspection of the results we decided to fix it at 87~deg. From the vertical \HI\ distribution analysis we obtained that on average $z_0$ is 1.1$\pm$0.4~kpc; therefore, we decided to fix this value in the model as well.
To check the validity of the fits, visual comparisons between the data and models were performed in many different representations of the cubes, such as channel map by channel map, the velocity field and the position velocity ($pv$) diagram parallel to the minor and major axis.
For the error bars on the final rotation curve, we used the difference between the approaching and receding side; a minimum realistic error of 2\kms\ was considered. Figure~\ref{fig:ff-rotation} shows that the rotation curve rises steeply in the innermost 4~kpc, then continues to rise slowly until the outermost radius. The maximum rotation speed is $\sim$72\kms. The de-projected \HI\ radial surface brightness profile derived by FAT shows the \HI\ surface density peak at $\Sigma_{\rm HI}\sim 5$~\Msun\,pc$^{-2}$ in the centre but is mostly constant around $\sim$5~\Msun\,pc$^{-2}$ (see Fig.~\ref{fig:ff-rotation}). The $pv$-diagram is shown in Fig.~\ref{fig:model+data}.

\section{Mass models}\label{sec:dm}

\subsection{Visible matter contribution}

The rotation velocities due to the gravitational potentials of the stellar and gaseous disks were determined separately using the task {\sc rotmod} of the Groningen Image Processing System \citep[GIPSY, ][]{1992ASPC...25..131V}. Since the old stellar population dominates the total stellar mass in late-type galaxies and mid-infrared emission is less susceptible to dust extinction and is also not affected by recent star formation, the stellar contribution ($V_*$) was derived using the stellar mass profile obtained by \cite{wang-2017}  from WISE 3.4\,$\mu$m infrared images. We assumed the vertical distribution as

\begin{displaymath}
D_*(z)= \frac{{\rm exp}(\frac{-z}{z_0})}{z_0},
\end{displaymath}

\noindent where $z_0$ is the disk scale-height ($z_0$=0.47~kpc), see for instance \cite{1981A&A....95..105V,1981A&A....95..116V}. The gaseous contribution ($V_{\rm gas}$) was derived using the de-projected \HI\  radial surface density profile as derived by FAT and TiRiFiC. We considered a vertical density distribution given by the exponential law
\begin{displaymath}
    D_{\rm gas}(z)=\frac{-{\rm exp}(\frac{z}{z_0})}{z_0},
\end{displaymath}
\noindent where the value of $z_0$=1.1$\pm$0.4~kpc (see Sec.~\ref{sec:thickness}). The \HI\ gas surface density distribution was then multiplied by a factor of 1.4 to account for the primordial helium, but we did not consider the presence of molecular $\rm H_2$ in our models; for late-type LSB spiral galaxies the ratio $\rm H_2/HI$ is $10^{-3}$ \citep{Matthews-2005}. \\

\subsection{Dark Matter Halo}

We used the so-called pseudo-isothermal halo (ISO) and the Navarro, Frenk \& White (NFW) density profiles to model the DM halo of the FF galaxy.

\begin{itemize}

\item The simplest model for a DM halo density profile is the pseudo-isothermal halo \citep{1991MNRAS.249..523B}. Its density profile is given by
\begin{displaymath}
  \rho_{\rm ISO}(R) = \frac{\rho_0}{1+(\frac{R}{R_c})^2},
\end{displaymath}

\begin{displaymath}
V_{\rm ISO}(R) = V_{\rm inf} \sqrt{4\phi G \rho_0 (R_C)^2 \left[1- \left[\frac{R_c}{R}arctan \left( \frac{R}{R_c}\right)\right]\right]},
\end{displaymath}

\noindent where $\rho_0$ is the central core density and $R_c$ is the core radius of the halo. The rotational velocity expression at any radius $R$, due to an ISO dark matter halo, is $V_{\rm ISO}$, and the asymptotic velocity of the halo is $V_{\rm inf}$

\item Based on numerical simulations of dark matter halos, Navarro, Frenk \& White \citep{1996ApJ...462..563N} described the radial density profile with the expressions:

\begin{displaymath}
\frac{\rho_{\rm NFW}}{\rho_{\rm crit}} = \frac{\delta_c}{\frac{R}{R_s} (1+\frac{R}{R_s})^2},
  \end{displaymath}

\begin{displaymath}
V_{NFW}(R)=V_{200} \sqrt {\frac{ ln(1+cx) - \frac{cx}{1+cx}}{x ln(1+c)-\frac{c}{1+c}}}.
  \end{displaymath}

where $R_s$ is the characteristic radius of the halo, $\rho_{\rm crit}$ is the critical density of the universe, $c$=$\frac{R_{200}}{Rs}$ is the concentration parameter and $x$=$\frac{R}{R_{200}}$. $R_{200}$ is the radius where the average density of the NFW halo is 200$\rho_{\rm crit}$. $R_{200}$ is in kpc and $V_{200}$=$0.73R_{200}$ is the rotation velocity at $R_{200}$ in km $\rm s^{-1}$.
\end{itemize}

Theoretically, since the FF galaxy is bulge-less, the net rotation velocity $V$ at a radius $R$ is obtained by adding in quadrature the rotational velocity due to the gravitational potential of the stars, gas and DM components:

\begin{displaymath}
V(R)^2 = \Upsilon_{\star} V_{\star}(R)^2 +  V_{\rm gas}(R)^2 + V_{\rm DM}(R)^2,
\end{displaymath}

\noindent where $\Upsilon_*$ is the stellar mass-to-light ratio.

We made use of {\sc rotmas}, a task in GIPSY \citep{gipsy} which allows interactive modelling of rotation curves, with options for the addition of dark matter halo contribution. We carried out the disk-halo decomposition using different assumptions of it: a maximum disk, a minimum disk with gas and a minimum disk.

\section{Mass-modelling: Results and discussion}

 As was already described, the rotation curve fitting procedure typically has as free parameters: the scale-length, the halo density and the mass-to-light ratio of the stellar component. The last value is not known a priory, its estimation is difficult, and may differ from galaxy to galaxy. \cite{cluver-2014} obtained an empirical relationship between mid-infrared emission of galactic disks and stellar mass-to-light ratios in the W1 and W2 WISE bands.Using this relation with the FF WISE photometric measurements available in \cite{wang-2017} we found that $\Upsilon_{\star}$=0.8.

As the approaching side of the galaxy is clearly kinetically disturbed, we performed the mass-modelling only considering the rotation curve derived from the receding half of the FF galaxy. 

\subsection{Maximum disk}

\begin{figure*}
\centering
\includegraphics[width=0.8\textwidth]{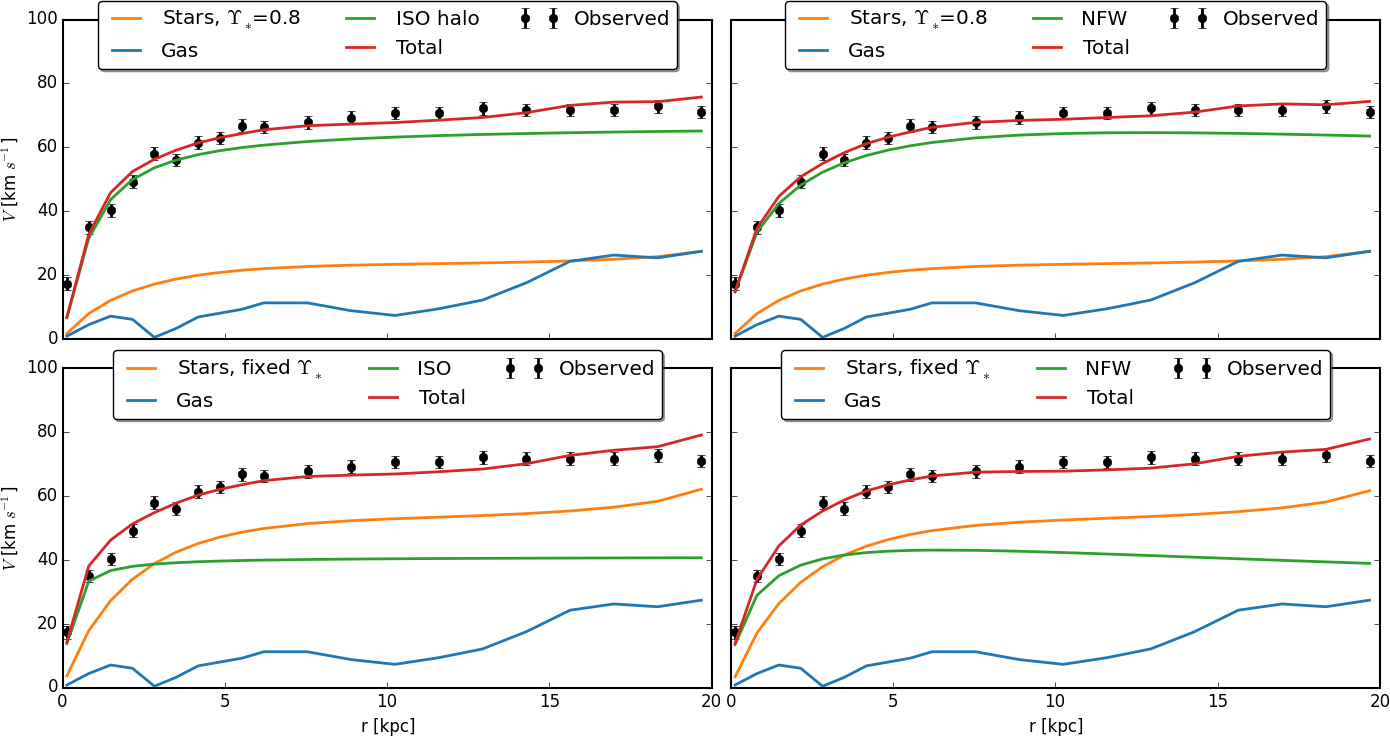}
\caption{Modelling \HI\ rotation curve of the Fourcade-Figueroa galaxy. {\bf Upper panels:} ISO and the NFW halo based mass model for the $Maximum$ $disk$ with $\Upsilon_{\star}$=0.8. {\bf Lower panels:} ISO and the NFW halo based mass model for the $Maximum$ $disk$ considering $\sim$75\% contribution to the peak of the rotation curve at $R = 2.2 r_{\rm s}$. The orange line indicates the rotation curve due to the stellar disk, the blue line due to the gas disk, the green line due to the dark matter halo and in red colour the best-fitting model rotation curve.}
\label{fig:dm}
\end{figure*}

In the maximum disk model, the stellar disk is scaled to the maximum possible value, dominating the underlying gravitational potential \citep{vanalbada-1985}. Thus, it returns an upper limit to the $\Upsilon_{\star}$ value and a lower bound on the DM distribution in the galaxy. The $maximum$ $disk$ model can be constructed in two different ways; by fixing $\Upsilon_{\star}$ to a maximum possible value, for FF we fixed it at 0.8, or by arbitrary fixing $\Upsilon_{\star}$ at the value which corresponds to the $\sim$75\% contribution to the peak of the rotation curve at $R = 2.2 r_{\rm s}$ \citep{sackett-1997}.
In Fig.~\ref{fig:dm} we show the rotation curve decomposition with the $maximum$ $disk$ model constructed considering both methods, and in Table~\ref{tab:fits} we list the results. The values of the mass-to-light ratio $\Upsilon_{\star}$ obtained for the best-fitting models considering the $\sim$75\% contribution to the peak of the rotation curve at $R = 2.2 r_{\rm s}$ are an order of magnitude higher than the value derived considering the stellar population. Both ISO and NFW halo give the higher values of reduced $\chi^2$ of all the fits. Previous studies on LSBs and some superthin galaxies indicate that most of these type of galaxies do not have a $maximum$ $disk$ \citep{deBlok-2008,banerjee-2017}, since the structure in the stellar disk is not seen in the rotation curve.

\subsection{Minimum disk}

\begin{figure*}
\centering
\includegraphics[width=0.8\textwidth]{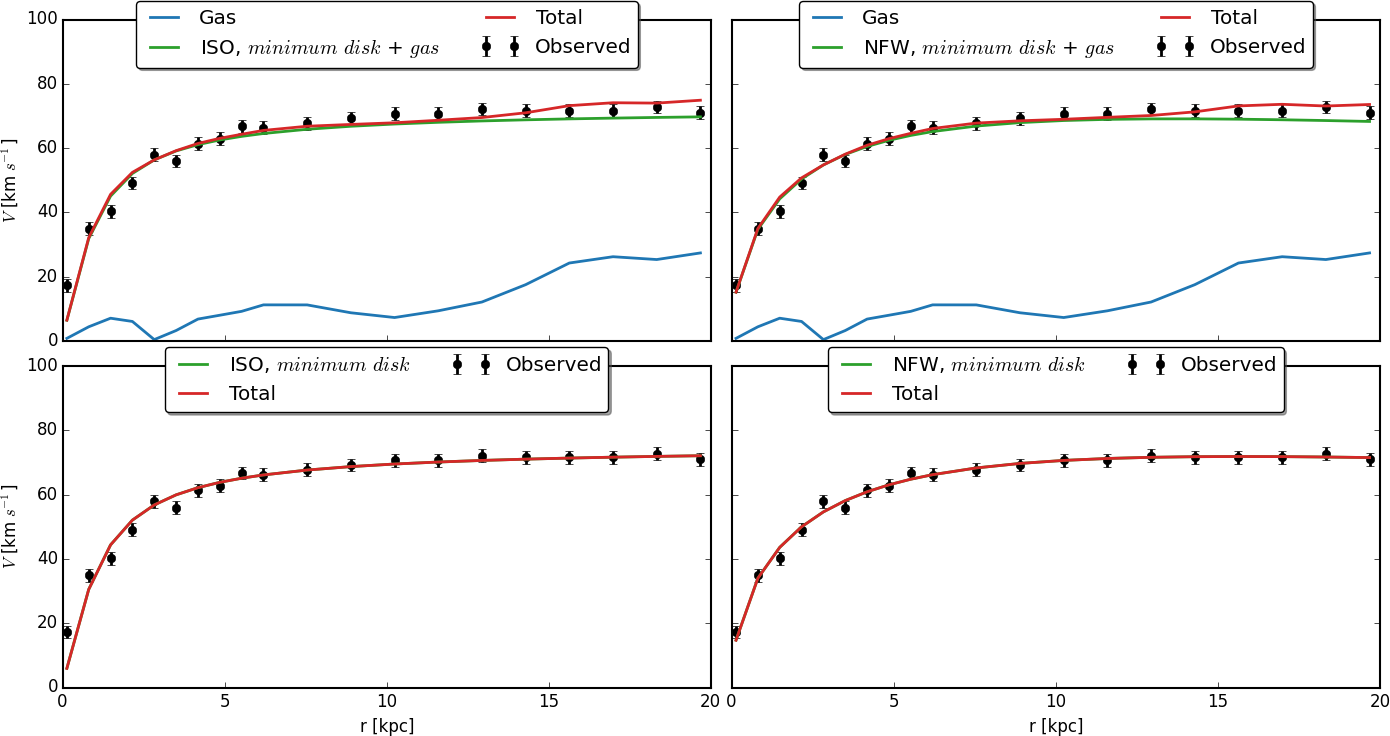}
\caption{Modelling \HI\ rotation curve of the Fourcade-Figueroa galaxy. {\bf Upper panels:} ISO and the NFW halo based mass model for the $minimum$ $disk$ $+$ $gas$. {\bf Lower panels:} ISO and the NFW halo based mass model for the FF galaxy for the $minimum$ $disk$. The orange line indicates the rotation curve due to the stellar disk, the blue line due to the gas disk, the green line due to the dark matter halo and in red colour the best-fitting model rotation curve.}
\label{fig:dm1}
\end{figure*}

The $minimum$ $disk$ model assumes that the rotation curve is entirely due to the presence of dark matter and therefore sets an upper bound on the dark matter density. We constructed two different $minimum$ $disk$ models: a $minimum$ $disk$ $with$ $gas$, this model only considers the contribution of the neutral hydrogen and helium to the rotation curve. 
In Fig.~\ref{fig:dm1}, upper panels, we show the rotation curve decomposition with the $minimum$ $disk$ + $gas$ and in Table~\ref{tab:fits} we list the results. And, a $minimum$ $disk$ where the stellar and gas disk's contribution are zero. In Fig.~\ref{fig:dm1}, lower panels, we show the rotation curve decomposition with the $minimum$ $disk$.\\

\begin{table*}
  \caption{Results of the mass modelling of the FF galaxy for ISO and NFW dark matter halo profiles.}
\centering
\begin{tabular}{lcccccccc}
\hline
Model      & $\Upsilon_{\star}$ & $\Upsilon_{\rm gas}$ & $R_C$ & $R_C/R_D$  &$\rho_0$ & $\chi^2$  \\
ISO                &       &          & [kpc] & &[$10^{-3}$ \Msun \, $\rm pc^{-3}$] &  \\
\hline
\hline
Maximum disk fixed $\Upsilon_*$ & 0.8  & 1.0 & 0.79$\pm$0.11 &0.2& 131$\pm$31 & 3.2 \\
Maximum disk     &    4.13        &   1.0        & 0.20$\pm$0.06   & 0.05 & 7165$\pm$405 & 3.0  \\
Minimum disk with gas &     0         & 1.0    & 0.8$\pm$0.1   & 0.13& 126$\pm$25 & 3.2\\
Minimum disk          &      0        &   0  & 0.9$\pm$0.1 &0.15& 109$\pm$18 &  2.7 \\
  \hline
Model                &  & & $c$ & $R_{200}$ &  $v_{200}$ &  \\
NFW                                &       &           &  & [kpc] & [\kkms]  &\\
                              
                                \hline
                                \hline
Maximum disk fixed $\Upsilon_*$ & 0.8 & 1.0 & 7.1$\pm$0.3 & 42$\pm$1 &$\sim$30 & 1.0\\
Maximum disk     &    4.13         &   1.0        &  8.6$\pm$1.1 &26$\pm$1 &$\sim$19 & 1.8\\
Minimum disk with gas &    0          & 1.0    &7.1$\pm$0.3 &44.8$\pm$0.6   & $\sim$32 &0.9 \\
Minimum disk          &   0          &  0   & 6.5$\pm$0.2   & 47.3$\pm$0.5 & $\sim$54 & 0.6\\

\hline
\end{tabular}
\label{tab:fits}
{\flushleft Notes: $R_c$ is the core radius of the ISO DM halo, $\rho_{0}$ is the central core density of the ISO DM halo, $c$ is the concentration parameter of the NFW DM halo, $R_{200}$ is the radius were the average density of the NFW DM halo is 200$\rho_{\rm crit}$ and parameters $\Upsilon_{\star}$ and $\Upsilon_{gas}$ are the scaling factor from the stellar and gaseous disk were fixed in order to improve the results.}
\end{table*}

We present the mass-modelling of the FF galaxy, carried out using ISO and NFW DM density profiles; the results are listed in table~\ref{tab:fits}. The best reduced $\chi^2$ values were thrown by the $minimum$ $disk$, especially the one obtained with the NFW DM halo. Thus, the contribution of the visible matter to the net gravitational potential is negligible. This result is in agreement with the trend observed in LSB and other superthin galaxies \citep{deBlok-2008,banerjee-2017,kurapati-2018}. We find that the ratio between the core radius of the halo and the scale-length of the optical disk ($R_C/R_D$) is always less than two for the ISO DM halo. This indicates that the core of the ISO halo is compact; consequently, the dark matter halo dominates at the inner radii as well. \cite{banerjee-2017} obtained similar results for 3 superthin galaxies (UGC\,7321, IC\,5249 and IC\,2233) as well as \cite{kurapati-2018} (FGC\,1540). The compactness of the dark matter halo may be responsible for the superthin disk distribution \citep{Banerjee-2013}.\\

\subsection{Mass-modelling with MOND formalism}

The Modified Newtonian Dynamics (MOND) postulates that Newtonian dynamics breaks down at small acceleration \citep{milgrom1983}. This formalism rules out the presence of dark matter and takes into account the self-gravity of the stars and gas in galactic dynamics studies. There are two free parameters: the stellar mass-to-light ratio ($\Upsilon_{\star}$) and the acceleration per unit length ($a_0$).

\begin{figure}
\centering
\includegraphics[width=0.45\textwidth]{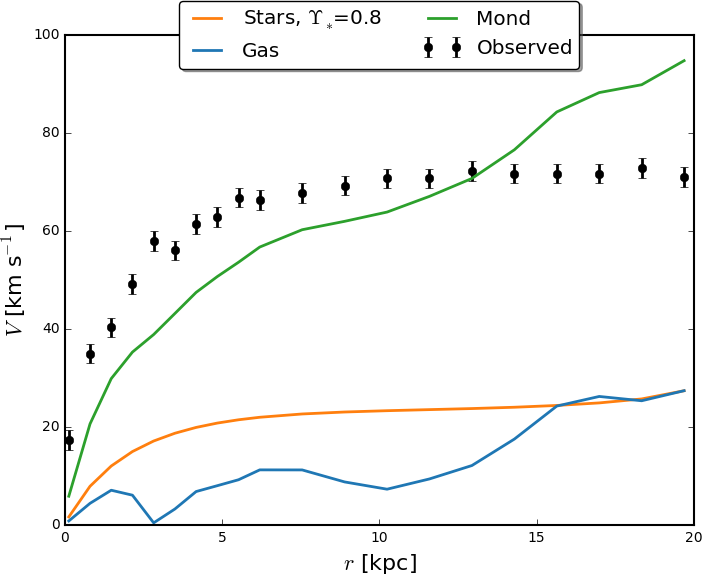}
\caption{Modelling \HI\ rotation curve of the Fourcade-Figueroa galaxy using MOND. The orange line indicates the rotation curve due to the stellar disk, the blue line the rotation curve due to the gas disk, and the green line is the MOND best-fitting model rotation curve.}
\label{fig:mond}
\end{figure}

In Fig.~\ref{fig:mond} we show the best-fitting rotation curve for the MOND formalism. In order to achieve a physically realistic value of $a_0$, $\Upsilon_{\star}$ was fixed to 0.8. The acceleration parameter resulted in $a_0=$2694$\pm$520~km$^2$\,s$^{-1}$\,kpc$^{-1}$. The reduced $\chi^2$ is 42; this means that in the case of FF the centrally concentrated mass indicated by the RC is not seen in the gas and stellar distribution. Which means there should be a very concentrated unseen baryonic component for MOND to work.

%-----------------------------------------------------------------

\section{The shred}

 Despite its superthin structure, the FF galaxy shows an asymmetry towards its north-west-side, clearly observed in both, optical and \HI\ moment maps. A series of features appears in the approaching side of the galaxy: 1) the iso-velocity contours start to exhibit distortions only a few kpc away from the centre; and, at the location of the kinematic anomalies, we can also see a thickening of the emission observed in the integrated moment map. However, due to the edge-on orientation of the galaxy, a more sophisticated analysis is required to determine at which radial location this occurs. And 2) the \HI\ gas steeply flares as the stellar component becomes fainter;  this is indicative of the mass in the disk changing \citep{sancisi-1979,Olling-1995}. Additionally, at the same location, the dispersion velocity is higher.
 
 The de-projected \HI\ peak surface density is $\Sigma_{\rm HI}\sim$5\Msun~pc$^{-2}$, close to the mean value proposed by \cite{Cayatte-1994} for Sbc galaxies ($\sim 6$ \Msun~pc$^{-2}$); this value is also in agreement with the results found by \cite{kurapati-2018} and \cite{banerjee-2017} for similar superthin galaxies. Furthermore, in most of the gas disk of FF, the gas surface density seems to lie below that required for efficient star formation \citep{kennicutt-1989}, and moreover, the overall density profile value is half of the peak surface density, see Fig.~\ref{fig:ff-rotation}. Although the galaxy is forming some stars, this result is consistent with the very low star formation rate derived from optical observations ($\log$(SFR)=$-1.0\rm\, M_{\odot}\,yr^{-1}$) and the low radio continuum emission previously reported \citep{Saponara-2012}; The evidence suggests that intense star formation activity is not the main cause of the observed disruption.
 
 In 1992, \cite{1992MNRAS.257..689T}, after the analysis of numerical simulations, proposed that the FF shred could be possibly originated because the FF galaxy undergoes a strong prograde interaction with the massive galaxy Cen~A. Currently, this theory is ruled out. A recent, more accurate, distance estimates put FF just beyond Cen~A \citep[$\sim$7~Mpc,][]{2015ApJ...805..144K}.
  Another possible explanation of the observed vertical thickness is the interaction with a companion dwarf galaxy, which enhance star formation in a small region in the western-most part of the disk, generating this and the higher dispersion observed. The available UV images from GALEX\footnote{\url{https://archive.stsci.edu/missions-and-data/galex}}, although they partially cover the galaxy, provide a hint of that.
 Despite the fact that the FOV is mostly filled with the galaxy, we visually looked for \HI\ low-mass companions of the FF galaxy, but no recognisable companions were observed. However, Cen\,A seems to have an overabundance of the faintest dwarfs in comparison to its simulated analogues \citep{Mueller-2019}, indicating that interactions are likely, and the scenario of a dwarf galaxy accreted like the Sag dSph by the Milky Way turns very probable \citep{RuizLara-2020}. 
 This would be contrary to superthins evolving in isolation. Of course, FF could be disturbed in such a way that it is evolving away from being a superthin galaxy although the quiescent and thin approaching side seems to counter this idea.  Up to now, most studies on superthin galaxies present this class as isolated objects, that is, devoid of interactions. An exception could be the galaxy  IC\,2233 for which the authors explore the possibility of interaction with smaller nearby galaxies \citep{uson-2003}.
 That interactions do not destroy the superthin disks in these galaxies fits with previous results \citep{obrien-2010,banerjee-2017,kurapati-2018} indicating that the compact DM halo is the main responsible for the superthin structure of the stellar disk, not isolation.  

\section{Summary}
We use the rotation curve, the \HI\ surface density profile along with the stellar profile obtained from the literature to construct the mass models for the FF galaxy. The FF rotation curve, as well as its de-projected \HI\ surface density profile, were derived by fitting a detailed 3D tilted ring model to the data. After the assumption of the ISO as well as the NFW dark matter halo, we found that both halos fit well the observed rotation curve, but the best reduced $\chi^2$ values were thrown by the $minimum$ $disk$ model considering the NFW halo. The results obtained from the mass-modelling implies that the dark matter halo is compact. We derived the thickness of the \HI\ disk as a function of the galactocentric radius and on the approaching side of the galaxy, the vertical \HI\ thickness shows an irregular profile. The FF galaxy shows an asymmetry towards its north-west-side, clearly observed at optical and \HI\ images. The new distance establishes FF away from Cen~A, which dismiss the hypothesis of past interaction between them. Even though, we did not find \HI\ low-mass companions for FF in our \HI\ cubes, tidal interaction could not be ruled out. Thus, the compact dark matter halo might be the main responsible for the superthin structure observed in the galaxy and not isolation.

\section*{Acknowledgements}

JS is grateful to Arunima Banerjee for use full discussions and to Jing Wang for the mass profile. This paper is based on observations obtained with the Australia Telescope Compact Array (ATCA) and the Giant Metrewave Radio Telescope (GMRT). ATCA is part of the Australia Telescope National Facility (ATNF) which is funded by the Australian Government for operation as a National Facility managed by CSIRO. We acknowledge the Gomeroi people as the traditional owners of the Observatory site. GMRT is operated by the National Centre for Radio Astrophysics of the Tata Institute of Fundamental Research. We thank the staff of both radio telescopes who made these observations possible. This research has made use of the NASA/IPAC extragalactic database (NED) that is operated by the Jet Propulsion Laboratory, California Institute of Technology, under contract with the National Aeronautics and Space Administration.
This work was partially supported by FCAG-UNLP, and ANPCyT project PICT 2017-0773.
PK is partially supported by the BMBF project 05A17PC2 for D-MeerKAT.

%%%%%%%%%%%%%%%%%%%% REFERENCES %%%%%%%%%%%%%%%%%%

\bibliographystyle{aa}
\bibliography{FF.bib} 

\begin{thebibliography}{63}
\expandafter\ifx\csname natexlab\endcsname\relax\def\natexlab#1{#1}\fi

\bibitem[{{Banerjee} \& {Bapat}(2017)}]{banerjee-2017}
{Banerjee}, A. \& {Bapat}, D. 2017, \mnras, 466, 3753

\bibitem[{{Banerjee} \& {Jog}(2013)}]{Banerjee-2013}
{Banerjee}, A. \& {Jog}, C.~J. 2013, \mnras, 431, 582

\bibitem[{{Banerjee} {et~al.}(2010){Banerjee}, {Matthews}, \&
  {Jog}}]{2010NewA...15...89B}
{Banerjee}, A., {Matthews}, L.~D., \& {Jog}, C.~J. 2010, \na, 15, 89

\bibitem[{{Begeman} {et~al.}(1991){Begeman}, {Broeils}, \&
  {Sanders}}]{1991MNRAS.249..523B}
{Begeman}, K.~G., {Broeils}, A.~H., \& {Sanders}, R.~H. 1991, \mnras, 249, 523

\bibitem[{{Cayatte} {et~al.}(1994){Cayatte}, {Kotanyi}, {Balkowski}, \& {van
  Gorkom}}]{Cayatte-1994}
{Cayatte}, V., {Kotanyi}, C., {Balkowski}, C., \& {van Gorkom}, J.~H. 1994,
  \aj, 107, 1003

\bibitem[{{Cluver} {et~al.}(2014){Cluver}, {Jarrett}, {Hopkins}, {Driver},
  {Liske}, {Gunawardhana}, {Taylor}, {Robotham}, {Alpaslan}, {Baldry}, {Brown},
  {Peacock}, {Popescu}, {Tuffs}, {Bauer}, {Bland-Hawthorn}, {Colless},
  {Holwerda}, {Lara-L{\'o}pez}, {Leschinski}, {L{\'o}pez-S{\'a}nchez},
  {Norberg}, {Owers}, {Wang}, \& {Wilkins}}]{cluver-2014}
{Cluver}, M.~E., {Jarrett}, T.~H., {Hopkins}, A.~M., {et~al.} 2014, \apj, 782,
  90

\bibitem[{{Colomb} {et~al.}(1984){Colomb}, {Loiseau}, \&
  {Testori}}]{1984ApL....24..139C}
{Colomb}, F.~R., {Loiseau}, N., \& {Testori}, J.~C. 1984, \aplett, 24, 139

\bibitem[{{Dalcanton} \& {Bernstein}(2000)}]{Dalcanton-2000}
{Dalcanton}, J.~J. \& {Bernstein}, R.~A. 2000, \aj, 120, 203

\bibitem[{{de Blok} \& {Bosma}(2002)}]{deBlock-2002}
{de Blok}, W.~J.~G. \& {Bosma}, A. 2002, \aap, 385, 816

\bibitem[{{de Blok} {et~al.}(2008){de Blok}, {Walter}, {Brinks},
  {Trachternach}, {Oh}, \& {Kennicutt}}]{deBlok-2008}
{de Blok}, W.~J.~G., {Walter}, F., {Brinks}, E., {et~al.} 2008, \aj, 136, 2648

\bibitem[{{de Vaucouleurs} {et~al.}(1991){de Vaucouleurs}, {de Vaucouleurs},
  {Corwin}, J., {Paturel}, \& {Fouque}}]{RC3}
{de Vaucouleurs}, G., {de Vaucouleurs}, A., {Corwin}, Jr., H.~G., {et~al.}
  1991, 1, Vol.~3, Third Reference Catalogue of Bright Galaxies (Berlin
  Heidelberg New York: Springer-Verlag)

\bibitem[{{Fourcade}(1970)}]{Fourcade}
{Fourcade}, C. 1970, Bol. Asoc. Argentina de Astron, 16, 10

\bibitem[{{Gerritsen} \& {de Blok}(1999)}]{gerritsen-1999}
{Gerritsen}, J. P.~E. \& {de Blok}, W.~J.~G. 1999, \aap, 342, 655

\bibitem[{{Gooch}(1996)}]{Gooch1996}
{Gooch}, R. 1996, in Astronomical Society of the Pacific Conference Series,
  Vol. 101, Astronomical Data Analysis Software and Systems V, ed. G.~H.
  {Jacoby} \& J.~{Barnes}, 80

\bibitem[{{Greisen}(2003)}]{Greisen2003}
{Greisen}, E.~W. 2003, {AIPS, the VLA, and the VLBA}, Vol. 285 (Astrophysics
  and Space Science Library), 109

\bibitem[{{Ho} {et~al.}(2011){Ho}, {Li}, {Barth}, {Seigar}, \&
  {Peng}}]{HO-2011}
{Ho}, L.~C., {Li}, Z.-Y., {Barth}, A.~J., {Seigar}, M.~S., \& {Peng}, C.~Y.
  2011, \apjs, 197, 21

\bibitem[{{J{\'o}zsa} {et~al.}(2012){J{\'o}zsa}, {Kenn}, {Oosterloo}, \&
  {Klein}}]{2012ascl.soft08008J}
{J{\'o}zsa}, G. I.~G., {Kenn}, F., {Oosterloo}, T.~A., \& {Klein}, U. 2012,
  {TiRiFiC: Tilted Ring Fitting Code}

\bibitem[{{Kamphuis} {et~al.}(2015){Kamphuis}, {J{\'o}zsa}, {Oh}, {Spekkens},
  {Urbancic}, {Serra}, {Koribalski}, \& {Dettmar}}]{2015ascl.soft07011K}
{Kamphuis}, P., {J{\'o}zsa}, G.~I.~G., {Oh}, S. .~H., {et~al.} 2015, {FAT:
  Fully Automated TiRiFiC}

\bibitem[{{Karachentsev} {et~al.}(1999){Karachentsev}, {Karachentseva},
  {Kudrya}, {Sharina}, \& {Parnovskij}}]{K1999}
{Karachentsev}, I.~D., {Karachentseva}, V.~E., {Kudrya}, Y.~N., {Sharina},
  M.~E., \& {Parnovskij}, S.~L. 1999, Bulletin of the Special Astrophysics
  Observatory, 47, 5

\bibitem[{{Karachentsev} {et~al.}(1993){Karachentsev}, {Karachentseva}, \&
  {Parnovskij}}]{K1993}
{Karachentsev}, I.~D., {Karachentseva}, V.~E., \& {Parnovskij}, S.~L. 1993,
  Astronomische Nachrichten, 314, 97

\bibitem[{Karachentsev \& Kudrya(2014)}]{Karachentsev_2014}
Karachentsev, I.~D. \& Kudrya, Y.~N. 2014, The Astronomical Journal, 148, 50

\bibitem[{{Karachentsev} {et~al.}(2015){Karachentsev}, {Tully}, {Makarova},
  {Makarov}, \& {Rizzi}}]{2015ApJ...805..144K}
{Karachentsev}, I.~D., {Tully}, R.~B., {Makarova}, L.~N., {Makarov}, D.~I., \&
  {Rizzi}, L. 2015, \apj, 805, 144

\bibitem[{{Kautsch}(2009)}]{Kautsch-2009}
{Kautsch}, S.~J. 2009, \pasp, 121, 1297

\bibitem[{{Kennicutt}(1989)}]{kennicutt-1989}
{Kennicutt}, Robert~C., J. 1989, \apj, 344, 685

\bibitem[{{Kennicutt} {et~al.}(2008){Kennicutt}, {Lee}, {Funes}, {J.}, {Sakai},
  \& {Akiyama}}]{kennicutt-2008}
{Kennicutt}, Robert~C., J., {Lee}, J.~C., {Funes}, J.~G., {et~al.} 2008, \apjs,
  178, 247

\bibitem[{{Koribalski} {et~al.}(2004){Koribalski}, {Staveley-Smith}, {Kilborn},
  {Ryder}, {Kraan-Korteweg}, {Ryan-Weber}, {Ekers}, {Jerjen}, {Henning},
  {Putman}, {Zwaan}, {de Blok}, {Calabretta}, {Disney}, {Minchin}, {Bhathal},
  {Boyce}, {Drinkwater}, {Freeman}, {Gibson}, {Green}, {Haynes}, {Juraszek},
  {Kesteven}, {Knezek}, {Mader}, {Marquarding}, {Meyer}, {Mould}, {Oosterloo},
  {O'Brien}, {Price}, {Sadler}, {Schr{\"o}der}, {Stewart}, {Stootman}, {Waugh},
  {Warren}, {Webster}, \& {Wright}}]{2004AJ....128...16K}
{Koribalski}, B.~S., {Staveley-Smith}, L., {Kilborn}, V.~A., {et~al.} 2004,
  \aj, 128, 16

\bibitem[{{Koribalski} {et~al.}(2018){Koribalski}, {Wang}, {Kamphuis},
  {Westmeier}, {Staveley-Smith}, {Oh}, {L{\'o}pez-S{\'a}nchez}, {Wong}, {Ott},
  {de Blok}, \& {Shao}}]{2018MNRAS.tmp..467K}
{Koribalski}, B.~S., {Wang}, J., {Kamphuis}, P., {et~al.} 2018, \mnras, 467

\bibitem[{{Kregel} {et~al.}(2002){Kregel}, {van der Kruit}, \& {de
  Grijs}}]{kregel-2002}
{Kregel}, M., {van der Kruit}, P.~C., \& {de Grijs}, R. 2002, \mnras, 334, 646

\bibitem[{{Kurapati} {et~al.}(2018){Kurapati}, {Banerjee}, {Chengalur},
  {Makarov}, {Borisov}, {Afanasiev}, \& {Antipova}}]{kurapati-2018}
{Kurapati}, S., {Banerjee}, A., {Chengalur}, J.~N., {et~al.} 2018, \mnras, 479,
  5686

\bibitem[{{Lauberts} \& {Valentijn}(1989)}]{1989spce.book.....L}
{Lauberts}, A. \& {Valentijn}, E.~A. 1989, {The surface photometry catalogue of
  the ESO-Uppsala galaxies} (Astrophysics and Space Science Library)

\bibitem[{{Matthews} {et~al.}(2005){Matthews}, {Gao}, {Uson}, \&
  {Combes}}]{Matthews-2005}
{Matthews}, L.~D., {Gao}, Y., {Uson}, J.~M., \& {Combes}, F. 2005, \aj, 129,
  1849

\bibitem[{{Matthews} \& {Wood}(2001)}]{Matthews-2001}
{Matthews}, L.~D. \& {Wood}, K. 2001, \apj, 548, 150

\bibitem[{{McGaugh} \& {de Blok}(1998)}]{1998ApJ...499...41M}
{McGaugh}, S.~S. \& {de Blok}, W.~J.~G. 1998, \apj, 499, 41

\bibitem[{{Milgrom}(1983)}]{milgrom1983}
{Milgrom}, M. 1983, \apj, 270, 365

\bibitem[{{Mosenkov} {et~al.}(2010){Mosenkov}, {Sotnikova}, \&
  {Reshetnikov}}]{mosenkov-2010}
{Mosenkov}, A.~V., {Sotnikova}, N.~Y., \& {Reshetnikov}, V.~P. 2010, \mnras,
  401, 559

\bibitem[{{M{\"u}ller} {et~al.}(2019){M{\"u}ller}, {Rejkuba}, {Pawlowski},
  {Ibata}, {Lelli}, {Hilker}, \& {Jerjen}}]{Mueller-2019}
{M{\"u}ller}, O., {Rejkuba}, M., {Pawlowski}, M.~S., {et~al.} 2019, \aap, 629,
  A18

\bibitem[{{Navarro} {et~al.}(1996){Navarro}, {Frenk}, \&
  {White}}]{1996ApJ...462..563N}
{Navarro}, J.~F., {Frenk}, C.~S., \& {White}, S. D.~M. 1996, \apj, 462, 563

\bibitem[{{O'Brien} {et~al.}(2010){O'Brien}, {Freeman}, \& {van der
  Kruit}}]{obrien-2010}
{O'Brien}, J.~C., {Freeman}, K.~C., \& {van der Kruit}, P.~C. 2010, \aap, 515,
  A63

\bibitem[{{Odewahn}(1994)}]{odewahn-1994}
{Odewahn}, S.~C. 1994, \aj, 107, 1320

\bibitem[{{Olling}(1995)}]{Olling-1995}
{Olling}, R.~P. 1995, \aj, 110, 591

\bibitem[{{Prasad} \& {Chengalur}(2011)}]{2011ascl.soft12007P}
{Prasad}, J. \& {Chengalur}, J. 2011, {FLAGCAL: FLAGging and CALlibration
  Pipeline for GMRT Data}, Astrophysics Source Code Library

\bibitem[{{Reshetnikov} \& {Combes}(1997)}]{reshetnikov-1997}
{Reshetnikov}, V. \& {Combes}, F. 1997, \aap, 324, 80

\bibitem[{{Roennback} \& {Bergvall}(1995)}]{Roennback-1995}
{Roennback}, J. \& {Bergvall}, N. 1995, \aap, 302, 353

\bibitem[{{Rubin} {et~al.}(1978){Rubin}, {Ford}, \&
  {Thonnard}}]{1978ApJ...225L.107R}
{Rubin}, V.~C., {Ford}, W.~K., J., \& {Thonnard}, N. 1978, \apj, 225, L107

\bibitem[{{Ruiz-Lara} {et~al.}(2020){Ruiz-Lara}, {Gallart}, {Bernard}, \&
  {Cassisi}}]{RuizLara-2020}
{Ruiz-Lara}, T., {Gallart}, C., {Bernard}, E.~J., \& {Cassisi}, S. 2020, Nature
  Astronomy, 4, 965

\bibitem[{{Sackett}(1997)}]{sackett-1997}
{Sackett}, P.~D. 1997, \apj, 483, 103

\bibitem[{{Sancisi} \& {Allen}(1979)}]{sancisi-1979}
{Sancisi}, R. \& {Allen}, R.~J. 1979, \aap, 74, 73

\bibitem[{{Saponara} {et~al.}(2012){Saponara}, {Lefranc}, {Benaglia},
  {Andruchow}, \& {Koribalski}}]{Saponara-2012}
{Saponara}, J., {Lefranc}, V., {Benaglia}, P., {Andruchow}, I., \&
  {Koribalski}, B. 2012, Boletin de la Asociacion Argentina de Astronomia La
  Plata Argentina, 55, 353

\bibitem[{{Sault} {et~al.}(1995){Sault}, {Teuben}, \& {Wright}}]{Sault1995}
{Sault}, R.~J., {Teuben}, P.~J., \& {Wright}, M.~C.~H. 1995, in Astronomical
  Society of the Pacific Conference Series, Vol.~77, Astronomical Data Analysis
  Software and Systems IV, ed. R.~A. {Shaw}, H.~E. {Payne}, \& J.~J.~E.
  {Hayes}, 433

\bibitem[{{Schlegel} {et~al.}(1998){Schlegel}, {Finkbeiner}, \&
  {Davis}}]{1998ApJ...500..525S}
{Schlegel}, D.~J., {Finkbeiner}, D.~P., \& {Davis}, M. 1998, \apj, 500, 525

\bibitem[{{Schwarzkopf} \& {Dettmar}(2001)}]{schwarzkopf-2001}
{Schwarzkopf}, U. \& {Dettmar}, R.~J. 2001, \aap, 373, 402

\bibitem[{{Serra} {et~al.}(2015){Serra}, {Westmeier}, {Giese}, {Jurek},
  {Fl{\"o}er}, {Popping}, {Winkel}, {van der Hulst}, {Meyer}, {Koribalski},
  {Staveley-Smith}, \& {Courtois}}]{Serra-2015}
{Serra}, P., {Westmeier}, T., {Giese}, N., {et~al.} 2015, \mnras, 448, 1922

\bibitem[{{Thomson}(1992)}]{1992MNRAS.257..689T}
{Thomson}, R.~C. 1992, \mnras, 257, 689

\bibitem[{{Toth} \& {Ostriker}(1992)}]{Toth-1992}
{Toth}, G. \& {Ostriker}, J.~P. 1992, \apj, 389, 5

\bibitem[{{Tully} {et~al.}(2013){Tully}, {Courtois}, {Dolphin}, {Fisher},
  {H{\'e}raudeau}, {Jacobs}, {Karachentsev}, {Makarov}, {Makarova},
  {Mitronova}, {Rizzi}, {Shaya}, {Sorce}, \& {Wu}}]{2013AJ....146...86T}
{Tully}, R.~B., {Courtois}, H.~M., {Dolphin}, A.~E., {et~al.} 2013, \aj, 146,
  86

\bibitem[{{Uson} \& {Matthews}(2003)}]{uson-2003}
{Uson}, J.~M. \& {Matthews}, L.~D. 2003, \aj, 125, 2455

\bibitem[{{van Albada} {et~al.}(1985){van Albada}, {Bahcall}, {Begeman}, \&
  {Sancisi}}]{vanalbada-1985}
{van Albada}, T.~S., {Bahcall}, J.~N., {Begeman}, K., \& {Sancisi}, R. 1985,
  \apj, 295, 305

\bibitem[{{van der Hulst} {et~al.}(1992{\natexlab{a}}){van der Hulst},
  {Terlouw}, {Begeman}, {Zwitser}, \& {Roelfsema}}]{1992ASPC...25..131V}
{van der Hulst}, J.~M., {Terlouw}, J.~P., {Begeman}, K.~G., {Zwitser}, W., \&
  {Roelfsema}, P.~R. 1992{\natexlab{a}}, in Astronomical Society of the Pacific
  Conference Series, Vol.~25, Astronomical Data Analysis Software and Systems
  I, ed. D.~M. {Worrall}, C.~{Biemesderfer}, \& J.~{Barnes}, 131

\bibitem[{{van der Hulst} {et~al.}(1992{\natexlab{b}}){van der Hulst},
  {Terlouw}, {Begeman}, {Zwitser}, \& {Roelfsema}}]{gipsy}
{van der Hulst}, J.~M., {Terlouw}, J.~P., {Begeman}, K.~G., {Zwitser}, W., \&
  {Roelfsema}, P.~R. 1992{\natexlab{b}}, in Astronomical Society of the Pacific
  Conference Series, Vol.~25, Astronomical Data Analysis Software and Systems
  I, ed. D.~M. {Worrall}, C.~{Biemesderfer}, \& J.~{Barnes}, 131

\bibitem[{{van der Kruit} \&
  {Searle}(1981{\natexlab{a}})}]{1981A&A....95..105V}
{van der Kruit}, P.~C. \& {Searle}, L. 1981{\natexlab{a}}, \aap, 95, 105

\bibitem[{{van der Kruit} \&
  {Searle}(1981{\natexlab{b}})}]{1981A&A....95..116V}
{van der Kruit}, P.~C. \& {Searle}, L. 1981{\natexlab{b}}, \aap, 95, 116

\bibitem[{{Wang} {et~al.}(2017){Wang}, {Koribalski}, {Jarrett}, {Kamphuis},
  {Li}, {Ho}, {Westmeier}, {Shao}, {Lagos}, {Wong}, {Serra}, {Staveley-Smith},
  {J{\'o}zsa}, {van der Hulst}, \& {L{\'o}pez-S{\'a}nchez}}]{wang-2017}
{Wang}, J., {Koribalski}, B.~S., {Jarrett}, T.~H., {et~al.} 2017, \mnras, 472,
  3029

\bibitem[{{Zasov} {et~al.}(1991){Zasov}, {Makarov}, \&
  {Mikhailova}}]{1991SvAL...17..374Z}
{Zasov}, A.~V., {Makarov}, D.~I., \& {Mikhailova}, E.~A. 1991, Soviet Astronomy
  Letters, 17, 374

\end{thebibliography}

\end{document}